# Photoassociation of ultracold LiRb* molecules: observation of high efficiency and unitarity-limited rate saturation

Sourav Dutta,[1,*] John Lorenz,[1] Adeel Altaf,[1] D. S. Elliott,[1,2] and Yong P. Chen[1,2,†]

[1] *Department of Physics, Purdue University, West Lafayette, IN 47907, USA*
[2] *School of Electrical and Computer Engineering, Purdue University, West Lafayette, IN 47907, USA*

We report the production of ultracold heteronuclear $^7$Li$^{85}$Rb molecules in excited electronic states by photoassociation (PA) of ultracold $^7$Li and $^{85}$Rb atoms. PA is performed in a dual-species $^7$Li-$^{85}$Rb magneto-optical trap (MOT) and the PA resonances are detected using trap loss spectroscopy. We identify several strong PA resonances below the Li $(2s\ ^2S_{1/2})$ + Rb $(5p\ ^2P_{3/2})$ asymptote and experimentally determine the long range $C_6$ dispersion coefficients. We find a molecule formation rate ($P_{LiRb}$) of $3.5\times10^7$ s$^{-1}$ and a PA rate coefficient ($K_{PA}$) of $1.3\times10^{-10}$ cm$^3$/s, the highest among heteronuclear bi-alkali molecules. At large PA laser intensity, we observe the saturation of the PA rate coefficient ($K_{PA}$) close to the theoretical value at the unitarity limit.

Heteronuclear polar molecules have recently attracted enormous attention [1-17] owing to their ground state having a large electric dipole moment [16]. The long range anisotropic dipole-dipole interaction in such systems is the basis for a variety of applications including quantum computing [13], precision measurements [14], ultracold chemistry [2] and quantum simulations [15]. Heteronuclear bi-alkali molecules (XY, where X and Y are two different alkali atom species), only a small subset of polar molecules, have received special attention mainly because the constituent alkali atoms are easy to laser cool and can be easily associated to form molecules at ultracold temperatures. The two primary methods for production of heteronuclear bi-alkali molecules have been magneto-association (MA), as in the case of KRb, NaK and NaLi [1-4], and photoassociation (PA), as in the case of LiCs, RbCs, NaCs, KRb and LiK [5-12]. Such molecules can be transferred to their absolute ground state where they have significant dipole moment, for example, by Stimulated Raman Adiabatic Passage (STIRAP) [1,12]. There is considerable interest in other heteronuclear combinations either due to their higher dipole moments, different quantum statistics or the possibility of finding simpler or more efficient methods for the production of ultracold molecules.

In this Rapid Communication, we report a highly efficient production of ultracold $^7$Li$^{85}$Rb molecules by PA. Prior to our work (also see [18]), LiRb was one of the few bi-alkali molecules yet to be produced at ultracold temperatures. There is considerable interest in LiRb because the rovibronic ground state LiRb molecule is predicted to have a relatively high dipole moment of 4.1 Debye (exceeded only by LiCs and NaCs) [16], which makes it a strong candidate for many of the applications mentioned above. It is also interesting to note that bosonic $^{85}$Rb, $^{87}$Rb and $^7$Li, and the fermionic $^6$Li are among the more commonly trapped alkali atomic species. This can make LiRb molecules readily available in both fermionic and bosonic forms (depending on the Li isotope chosen), broadening the range of physics that can be studied. We provide the first step towards the production of such ultracold LiRb molecules. In our experiment, the $^7$Li$^{85}$Rb molecules are created in excited electronic states (denoted by LiRb*) by photoassociating ultracold $^7$Li and $^{85}$Rb atoms held in a dual-species magneto-optical trap (MOT). Contrary to previous expectation [19], we find very high LiRb* formation rate ($P_{LiRb}$) of $3.5\times10^7$ s$^{-1}$ which is promising for ultimately creating ultracold LiRb molecules in their rovibronic ground state. We also report a PA rate coefficient ($K_{PA}$) of $1.3\times10^{-10}$ cm$^3$/s, the highest among heteronuclear bi-alkali molecules. In addition, we observe the unitarity limited saturation of $K_{PA}$ at a value that is in agreement with the theory of Bohn and Julienne [20]. This is the first time that the saturation of $K_{PA}$ at the unitarity limit has been observed for heteronuclear bi-alkali polar molecules. This has important implications, for example, in the observation of atom-molecule oscillations and coherent control (which often requires understanding the strongly driven regime) [20-25].

Our experiments are performed in a dual-species MOT. The details of the apparatus have been described elsewhere [26]. We use a conventional MOT for $^7$Li and a dark MOT for $^{85}$Rb which allows us to trap a large number of atoms with minimal losses from light-assisted interspecies collisions [26]. The spatial overlap of the two MOTs is monitored using a pair of CCD cameras placed orthogonal to each other and a good overlap is ensured for the PA experiments. The $^{85}$Rb dark-MOT typically contains $N_{Rb} \sim 1\times10^8$ $^{85}$Rb atoms at a density ($n_{Rb}$) $\sim 4\times10^9$ cm$^{-3}$, with a majority of the $^{85}$Rb atoms in the lower ($F = 2$) hyperfine level of the $5s\ ^2S_{1/2}$ state. The $^7$Li MOT typically contains $N_{Li} \sim 6\times10^7$ $^7$Li atoms at a density ($n_{Li}$) $\sim 5\times10^9$ cm$^{-3}$, with a majority of the atoms in the upper ($F = 2$) hyperfine level of the $2s\ ^2S_{1/2}$ state. The atoms collide mainly along the $^7$Li$(2s\ ^2S_{1/2}, F = 2) + ^{85}$Rb$(5s\ ^2S_{1/2}, F = 2)$ channel. The fluorescence from both the MOTs is collected using a lens, separated using a dichroic mirror and recorded on two separate photodiodes.

Light for the PA measurements is produced by a Ti:Sapphire laser with a linewidth below 1 MHz, maximum output power of 480 mW and maximum mode-hop-free scan of 20 GHz. The PA laser beam is collimated to a $1/e^2$ diameter of 0.85 mm, leading to a maximum available average intensity of about 85 W/cm$^2$. In this article, we report PA resonances below the Li$(2s\ ^2S_{1/2})$+Rb$(5p\ ^2P_{3/2})$ asymptote i.e. near the D$_2$ line of Rb at 780 nm, while we have also recently observed PA resonances near the D$_1$ line of Rb at 795 nm [18]. PA resonances lead to the formation of LiRb* molecules which either spontaneously decay to LiRb molecules in the electronic ground state or to free Li and Rb atoms with high kinetic energies. Both mechanisms result in loss of the Li and Rb atoms from the MOT leading to a decrease in the MOT fluorescence. The formation of LiRb* molecules can thus be detected from this so called trap loss spectrum. In this work, the signature of LiRb* PA resonances is detected through the trap loss in the Li MOT (the trap loss signal of the Rb MOT is complicated due to the presence of numerous Rb$_2$ PA resonances in addition to the LiRb PA resonances). We have also verified that the observed LiRb PA resonances are absent unless both the Li and Rb MOTs are simultaneously present.

A part of the experimentally observed PA spectrum below the Li$(2s\ ^2S_{1/2})$+Rb$(5p\ ^2P_{3/2})$ asymptote is shown in Fig. 1. The detuning $\Delta_{PA}$ is measured with respect to the frequency $\nu_{res}(=384232.157\ \text{GHz})$ of the Rb$(5s\ ^2S_{1/2}, F=2)\rightarrow$ Rb$(5p\ ^2P_{3/2}, F'=3)$ transition, i.e. $\Delta_{PA}=\nu_{PA}-\nu_{res}$, where $\nu_{PA}$ is the frequency of the PA laser. The detuning $\Delta_{PA}$ is thus a measure of the binding energy ($E_B$) of the respective vibrational levels. For the spectrum shown in Fig. 1(a), the frequency of the PA laser was scanned at 10 MHz/s (while the MOT loading time is ~ 5s) and the intensity of the PA beam was 70 W/cm$^2$. The full spectrum in Fig. 1(a) was obtained by stitching together many short 4 GHz scans. The absolute frequencies of the PA lines are accurate to ± 100 MHz, although the frequency resolution is much higher and hence the relative frequency measurements are more accurate. The narrowest PA lines have a linewidth of ~ 33 MHz and we are able to resolve finer structures in the PA lines as shown in Fig. 1(b) (see supplementary information S1 for zoomed-in view of all observed PA resonances). Figure 1(c) shows the strongest PA line observed which corresponds to a LiRb* molecule production rate of 3.5×10$^7$ s$^{-1}$. We will elaborate on this particular PA line below but first discuss the assignment of the observed PA lines.

The theoretical potential energy curves at large internuclear separations ($R$) depend on the $C_6$ coefficients [27-29], an example of which is shown for LiRb in Fig. 2(a). In Hund's case (c), there are five electronic states asymptotic to the Li$(2s\ ^2S_{1/2})$+Rb$(5p\ ^2P_{3/2})$ asymptote [27-32]. These are labeled $n(\Omega^\sigma)$, where $\Omega$ is the projection of the total electronic angular momentum on the internuclear axis, σ = -/+ (only for Ω = 0) depending on whether or not the electronic wave function changes sign upon reflection at any plane containing the internuclear axis, and $n$ is a number denoting the $n^{\text{th}}$ electronic state of a particular $\Omega^\sigma$. The 3(0$^+$), 3(0$^-$) and 3(1) have very similar $C_6$ coefficients, are almost indistinguishable and form a triad. The 4(1) and 1(2) have very similar $C_6$ coefficients, are almost indistinguishable and form a diad. Only these five states are relevant for PA near the D$_2$ asymptote.

We have identified several vibrational levels belonging to the 3(0$^+$), 4(1) and 1(2) electronic states but did not observe the 3(0$^-$) and 3(1) states. For each vibrational level, we observe at most two lines corresponding to two rotational levels ($J$). While it is true that $s$-wave collisions dominate in our system since the $p$-wave centrifugal barrier (~ 1.9 mK, estimated using $C_6$ ~ 2500 atomic units for the Li$(2s\ ^2S_{1/2})$+Rb$(5s\ ^2S_{1/2})$ asymptote) is higher than the MOT temperatures (~1 mK and ~200μK for the Li and Rb MOTs, respectively), it does not provide a satisfactory explanation of why only up to two rotational levels are observed.

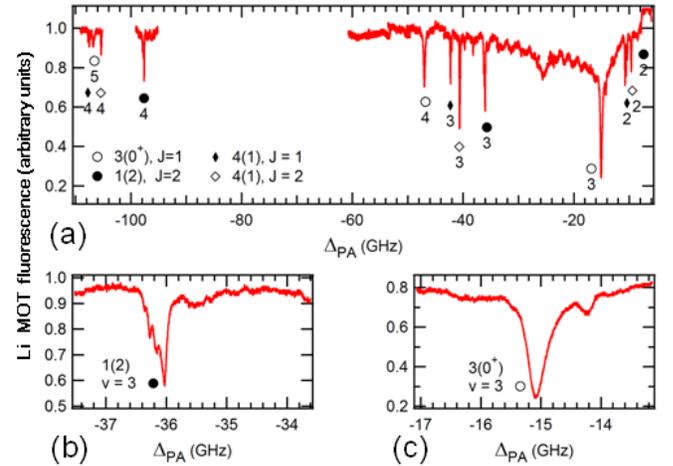

FIG. 1. (Color online) (a) PA spectra of LiRb obtained using trap loss spectroscopy. A reduction in the fluorescence of the Li MOT is observed whenever the PA laser is tuned through a LiRb* PA resonance. The open circles, filled circles, filled rhombus and open rhombus indicate lines belonging to the 3(0$^+$), 1(2), 4(1) $J$ =1 and 4(1) $J$ = 2 states, respectively, while the numbers indicate the vibrational level $v$ measured from the dissociation limit (see text for details). (b) PA spectrum near the $v$ = 3 level of the 1(2) state showing resolved hyperfine structure. (c) PA spectrum near the $v$ = 3 level of the 3(0$^+$) state which is broad and leads to a 70% trap loss.

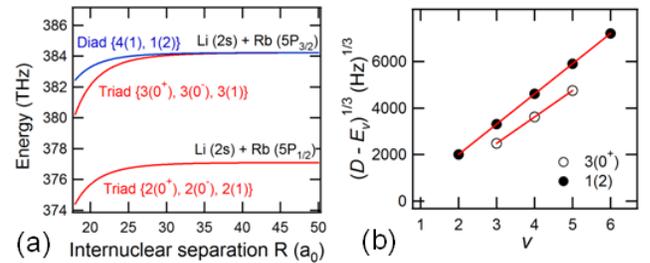

FIG. 2. (Color online) (a) The theoretical potential energy curves [27] for LiRb at large internuclear separations. Only curves near the Li (2$s$) + Rb (5$p$) asymptote are shown. (b) The fit to the LRB formula, Eq. (1), that is used to extract the $C_6$ coefficients from the experimentally measured PA line positions (to avoid clutter, only two representative states are shown).

The selection rules for the electric dipole transitions are: $\Delta\Omega = 0, \pm 1$, $\Delta J = 0, \pm 1$, $0^+\leftrightarrow 0^+$, $0^-\leftrightarrow 0^-$, $J \geq \Omega$, $\Delta J = 0$ is not allowed if $\Omega = 0$ for both states and $J = 0 \rightarrow J = 0$ transitions are not allowed [10, 30-32]. The Li and Rb atoms in the ground state collide along the $1(0^+)$, $1(0^-)$ or the $1(1)$ channels. Along with the assumption that only *s*-wave ($l = 0$) collisions contribute, this leads to the following allowed levels of the photoassociated LiRb* molecule: $0^+$ ($J = 0, 1, 2$), $0^-$ ($J = 0, 1, 2$), 1 ($J = 1, 2$) and 2 ($J = 2$). However, for reasons that remain to be understood, we observe at most two of the allowed rotational levels. For the $\Omega = 0^+$ state, we observe only one *J* and hence cannot accurately assign the value of *J*. On the other hand, the width and the hyperfine structure of the observed PA lines allow us to make an educated guess for *J* and also of $\Omega$. We assign the relatively wide (~ 250 MHz) PA lines to $3(0^+)$, $J = 1$ since (*i*) the linewidth of $3(0^+)$ resonances are expected to be large since the $3(0^+)$ state, which correlates to the $b\,^3\Pi$ - $A\,^1\Sigma^+$ complex at smaller internuclear distances, could undergo predissociation [32] due to the avoided crossing [33] between the $b\,^3\Pi$ and $A\,^1\Sigma^+$ states, (*ii*) they are expected to have no hyperfine structure, and (*iii*) they fit the expected line positions of the triad potential quite well as discussed in details below. We assign the lines with multiple resolved hyperfine splitting to $1(2)$, $J = 2$ since (*i*) the hyperfine splitting is expected to be the largest for these lines [34] and (*ii*) they fit the diad potential quite well. We assign the other lines to the $J = 1$ and $J = 2$ levels of the $4(1)$ state based on the agreement with the expected rotational constant.

For the $3(0^+)$ state, we prefer assigning $J = 1$ instead of $J = 0$ or $J = 2$ because there are at least twice as many allowed PA transitions from the $1(0^+)$, $1(0^-)$ and $1(1)$ collision channels that lead to the formation of $J = 1$ LiRb* molecules compared to $J = 0$ or $J = 2$ LiRb* molecules. In table 1 we report all the observed PA lines along with their assignments whenever possible. For levels with hyperfine structure, the position of the strongest line is reported. We note that the identification of the diad and triad potentials and the derived $C_6$ coefficients are expected to be quite accurate despite some uncertainties in the assignment of the angular momentum quantum numbers such as *J*.

Table 1. The values of $-\Delta_{PA}$ (in GHz) for which PA lines are observed.

| State | $v = 2$ | $v = 3$ | $v = 4$ | $v = 5$ | $v = 6$ |
|---|---|---|---|---|---|
| $3(0^+), J = 1$ | | 15.08 | 47.03 | 106.76 | |
| $1(2), J = 2$ | 7.91 | 36.06 | 97.61 | 205.52 | 373.07 |
| $4(1), J = 2$ | 9.62 | 40.69 | 105.36 | | |
| $4(1), J = 1$ | 10.75 | 42.35 | 107.50 | | |

Table 2. The values of $C_6$ coefficients (in atomic units) for the Li ($2s\ ^2S_{1/2}$) + Rb ($5p\ ^2P_{3/2}$) asymptote measured experimentally in this work, and a comparison with three different theoretical predictions. The numbers in parentheses indicate the uncertainties in the experimental determination. The experimentally determined values of $v_D$ are also included.

| | This work | | [27] | [28] | [29] |
|---|---|---|---|---|---|
| | $v_D$ | $C_6$ | $C_6$ | $C_6$ | $C_6$ |
| $3(0^+)$ | 0.80 | 20160 (950) | 20670 | 24980 | 26744 |
| $1(2)$ | 0.40 | 9235 (490) | 9205 | 11308 | 9431 |
| $4(1)$ | 0.24 | 10190 (420) | 9205 | 11308 | 9431 |

To extract the $C_6$ coefficients we use the LeRoy-Bernstein (LRB) formula [35]:

$$D - E_v = A_6(v - v_D)^3 \qquad (1)$$

where *v* is the vibrational quantum number measured from the dissociation limit (i.e. measured such that $v = 1$ is the least bound state), $v_D$ ($0 < v_D < 1$) is the vibrational quantum number at dissociation, $-(D - E_v)$ is the (negative) binding energy $E_B$, $D$ is the (positive) dissociation energy, $E_v$ is the (positive) energy of the $v^{\text{th}}$ vibrational level and $A_6 = 16\sqrt{2}\pi^3\hbar^3 / \{[B(2/3, 1/2)]^3 \mu^{3/2} C_6^{1/2}\}$, with $\mu$ being the reduced mass of $^7$Li$^{85}$Rb and $B$ being the Beta function [$B(2/3, 1/2) = 2.5871$]. Note that $(D - E_v) - E_{rot} = -h\Delta_{PA}$ is experimentally measured, where $E_{rot} = B_v[J(J+1) - \Omega^2]$ is the rotational energy, and $B_v$ is the rotational constant.

For each $\Omega$ state, we first assign the vibrational quantum number (*v*) and then plot $(-h\Delta_{PA})^{1/3} \approx (D - E_v)^{1/3}$ against *v*, neglecting the small contribution from $E_{rot}$. We then derive the values of $v_D$ and $A_6$ (and hence $C_6$) from a fit to Eq. (1), an example of which is shown in Fig. 2(b). Accounting for $E_{rot}$, which is small ($135 \le B_v \le 515$ MHz), does not change these values significantly (also see supplementary information S2) and the $C_6$ coefficients are within the quoted uncertainties even if the rotational assignment (*J*) is changed by $\pm 1$. We compare the experimentally determined $C_6$ coefficients with the theoretical values available (Table 2) and generally find good agreement with Ref. [27] and reasonable agreement with others.

We now turn our attention to the strongest PA lines at $\Delta_{PA} = -15.08$ GHz ($v_{PA} = 384217.07$ GHz) for which we observe 70% losses in the Li MOT. In Fig. 3(a) we show the time evolution of the number of atoms in the Li MOT when the resonant PA beam, with average intensity = 74 W/cm$^2$, is turned on at $t = 0$ (also see supplementary information S3). From the initial slope of the curve, we determine that $3.5 \times 10^7$ Li atoms are lost per second. This implies a LiRb* production rate ($P_{LiRb}$) of $3.5 \times 10^7$ s$^{-1}$, which is among the highest observed for heteronuclear bi-alkali molecules. Since the Rb MOT is larger than the Li MOT, the PA rate coefficient $K_{PA}$ is simply given by: $K_{PA} = P_{LiRb}/(n_{Rb}N_{Li})$. Using the typical values of $N_{Li}$ and $n_{Rb}$, the maximum value that we obtain is $K_{PA} = 1.3 \times 10^{-10}$ cm$^3$/s. The value of $K_{PA}$ is accurate to within 50%, the major uncertainty coming from the determination of $n_{Rb}$ (note that the uncertainties in the measurement of $N_{Li}$ cancel out). We also observe the saturation of $P_{LiRb}$ and hence $K_{PA}$, for PA laser intensities exceeding 60 W/cm$^2$ (Fig. 3(b)). The saturated value of $K_{PA}$ is estimated to be around $1.3(7) \times 10^{-10}$ cm$^3$/s. This value can be compared to the predicted theoretical value at the unitarity limit where the scattering matrix element becomes unity [20,24,31]:

$K_{PA,\text{unitarity}} = \pi v_{rel}/k^2 = \hbar^2 \sqrt{2\pi/\mu^3 k_B T} = 2.1 \times 10^{-10}$ cm$^3$/s,

where $v_{rel} = \sqrt{8k_B T/\pi\mu}$ is average relative velocity of the atoms, $k = \sqrt{2\mu k_B T/\hbar^2}$, $\mu$ is the reduced mass and $T = 1$ mK is the temperature of the Li atoms. Given the uncertainty in the values of $n_{Rb}$ and $T$, we consider that the agreement between experiment and theory is fairly good. We note that we performed a similar analysis for PA of $^{85}$Rb$_2$ we measured and found good agreement with theory [20] and with previous experimental reports.

This is the first time that unitarity limited saturation of $K_{PA}$ has been observed for heternuclear polar molecules. In Fig. 3(c) we plot the maximum observed $K_{PA}$ values for different polar molecules along with the theoretical values at the unitarity limit [20]. It is seen that the experimentally observed $K_{PA}$ value for LiRb is higher than all other species so far and approaches the unitarity limit. We note that $K_{PA}$ can be further increased by lowering $T$. It is also seen that for most other species the experimental maximum values differ substantially from the values at the unitarity limit. Moreover, the measured $K_{PA}$ values for LiCs [36] and NaCs [37] have large error bars and are only accurate within a factor of 10. The error bars are large in those measurements because the measurements were performed using Resonance Enhanced Multi-Photon Ionization (REMPI) for which it is difficult to calibrate the ionization, ion collection and detection efficiencies. PA induced trap loss was used to measure PA rates in RbCs [9] and LiK [10] but the rates were significantly lower. In this regard, LiRb is a welcome exception with very high PA rate and the observation of PA-induced trap loss allows relatively precise determination of the PA rate coefficient $K_{PA}$. We note that high PA rate (though lower than that observed here) is also observed for PA near the $\text{Li}(2s\ ^2S_{1/2})+\text{Rb}(5p\ ^2P_{1/2})$ asymptote [18].

We also note that we observe a very high LiRb* molecule production rate ($P_{LiRb}$) of $3.5\times10^7$ s$^{-1}$. Assuming that a small fraction of the LiRb* molecules would spontaneously decay to electronic ground state LiRb molecules, this amounts to a substantial production rate of LiRb molecules in the electronic ground state (interested readers are directed to supplementary information S4 for discussions on the prospects and schemes to create ground state LiRb molecules). Detection of such ground state molecules using REMPI, guided by our previous spectroscopy work [38], is currently being pursued.

In conclusion, we have produced ultracold LiRb* molecules by photoassociation and derived the $C_6$ dispersion coefficients for the $\text{Li}(2s\ ^2S_{1/2})+\text{Rb}(5p\ ^2P_{3/2})$ asymptote. We find unexpectedly high LiRb* molecule production rate of $3.5\times10^7$ s$^{-1}$. We observe a PA rate coefficient ($K_{PA}$) of $1.3\times10^{-10}$ cm$^3$/s, highest among heteronuclear bi-alkali molecules, and report the PA laser intensity driven saturation of $K_{PA}$ at a value close to the unitarity limit. We note that PA could lead to formation of LiRb molecules in the deeply bound vibrational levels of the ground electronic $X\ ^1\Sigma^+$ state. These can then be transferred to the $X\ ^1\Sigma^+$ ($v'' = 0$) state by techniques such as STIRAP [1,12] or optical pumping [39].

*Acknowledgments*: We thank Olivier Dulieu, Jesús Pérez-Ríos and Michael Bellos for helpful discussions and feedback. S.D. acknowledges financial support from Purdue University in the form of the Bilsland Dissertation Fellowship. Support during early stages of this work by the NSF (CCF-0829918), and through an equipment grant from the ARO (W911NF-10-1-0243) are gratefully acknowledged.

**Supplementary information**

S1. Zoomed-in view of all observed PA lines

PA spectra for all the observed lines (in the order of decreasing frequency) are provided below. The detuning $\Delta_{PA}$ is measured with respect to the frequency $\nu_{res}$ (= 384232.157 GHz) of the Rb ($5s\ ^2S_{1/2}$, $F = 2$) → Rb ($5p\ ^2P_{3/2}$, $F' = 3$) transition, i.e. $\Delta_{PA} = \nu_{PA} - \nu_{res}$, where $\nu_{PA}$ is the frequency of the PA laser. Unlike the figures in the main manuscript, these scans were taken at different scan speeds and PA intensities and hence a direct comparison of the line intensities should not be made. The open circles, filled circles, filled rhombus and open rhombus indicate lines belonging to the 3(0$^+$), 1(2), 4(1) $J$ =1 and 4(1) $J$ = 2 states respectively, while the numbers indicate the vibrational quantum number $v$ measured from the dissociation limit. Also of future interest are the resolved hyperfine structures.

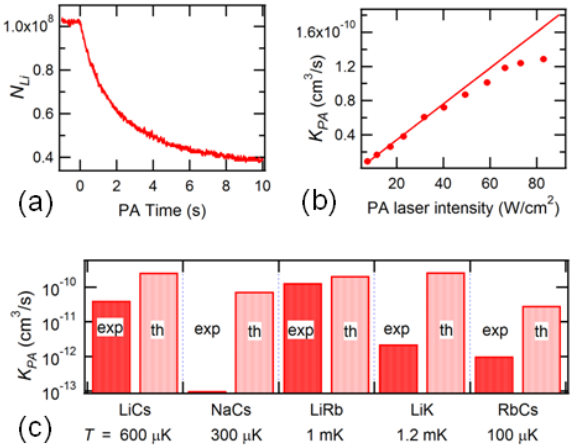

FIG. 3. (Color online) (a) The evolution of the atom number $N_{Li}$ in the Li MOT when the on resonance PA light ($\nu_{PA}$ = 384217.07 GHz, the resonance shown in Fig. 1c) is turned on at $t = 0$. The LiRb* production rate ($P_{LiRb}$) is estimated from the slope near $t = 0$. (b) The PA rate coefficient ($K_{PA}$) as a function of the average PA laser intensity. The rate starts to saturate beyond 60 W/cm$^2$. The solid line is a linear fit of $K_{PA}$ in the low-intensity regime with a slope $2\times10^{-12}$ (cm$^3$/s)/(W/cm$^2$). (c) Comparison of PA rate coefficient $K_{PA}$ of different polar molecules. The left (dark color, exp) bars denote the maximum experimentally observed values of $K_{PA}$ while the right (light color, th) bars denote the theoretical values at the unitarity limit at the temperatures indicated. Even at a relatively high temperature, the $K_{PA}$ for LiRb is the highest and is close to the theoretical prediction of the unitarity limit.

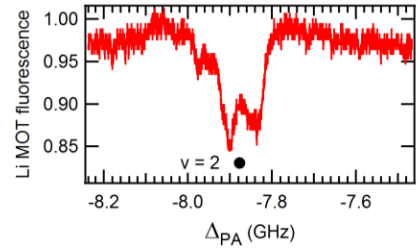

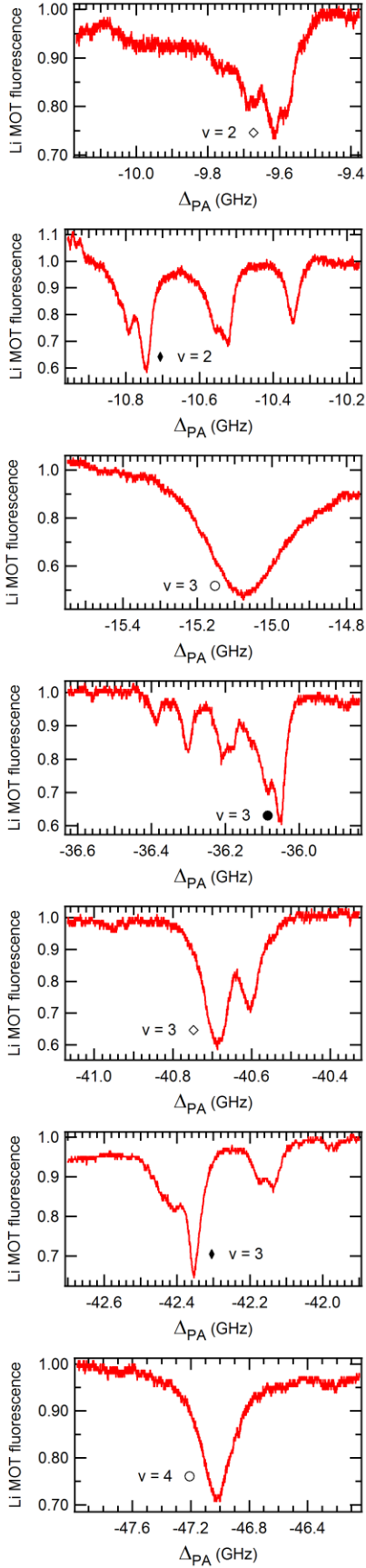

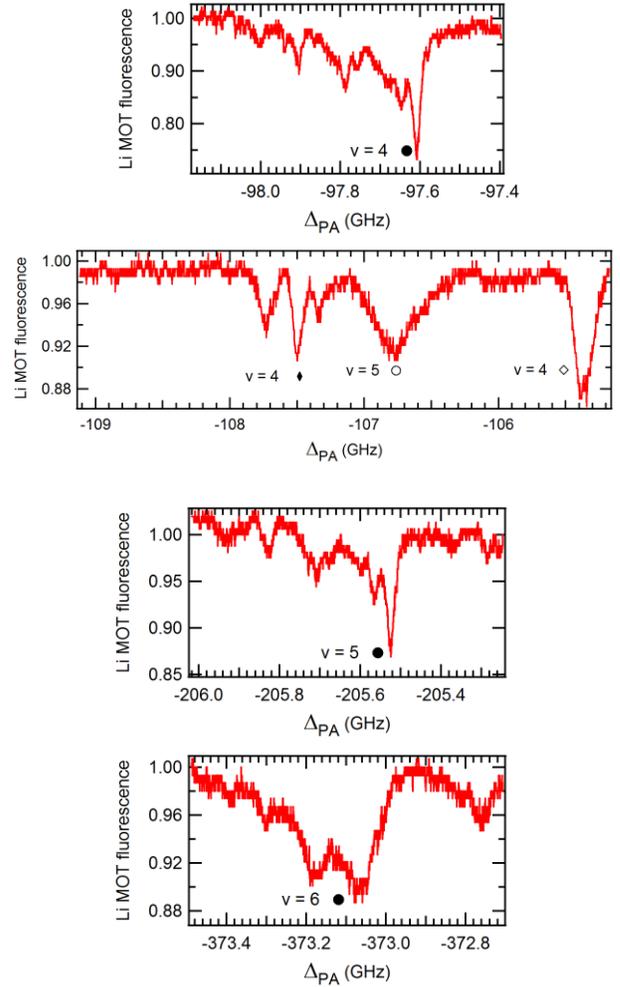

## S2. Discussion on the small contribution from the rotational energy $E_{rot}$

In order to reduce the uncertainties in the $C_6$ coefficients due to the possible contribution from $E_{rot}$, we do the following for the electronic states in which only one rotational level is observed. Using the $C_6$ coefficients derived in the main manuscript we first calculate the potential $V(R) = -C_6/R^6$. We then equate the measured $E_B$ of the $v^{th}$ vibrational level to $V(R)$ to derive the outer turning point $R_{out}$ of the level which then gives the rotational constant $B_v \approx \hbar^2/2\mu R_{out}^2$ of the $v^{th}$ vibrational level. We then subtract out the rotational energy and plot $(-h\Delta_{PA} + E_{rot})^{1/3}$ vs. $v$ again and derive a new set of the values of $v_D$ and $A_6$ (and hence $C_6$) from a fit to Eq. (1). After this iteration, the new values of $v_D$ and $C_6$ are very close to the initial values. The values of $v_D$ and $C_6$ coefficients are reported in Table 2 of the main manuscript. We note that these values do not vary significantly (i.e. are within the quoted uncertainties) even if the rotational assignment ($J$) is changed by ±1.

$$D - E_v = A_6(v - v_D)^3 \qquad (1)$$

## S3. Saturation of PA rate

The figure below shows the time evolution of the Li MOT atom number when the photoassociation (PA) laser ($\nu_{PA} = 384217.07$ GHz) is turned on at $t = 0$ s. The time evolution for different average PA laser intensities (I) is shown. The saturation of molecule formation rate $P_{LiRb}$, and hence the PA rate $K_{PA}$, is evident for intensities above 60 W/cm$^2$. We did not observe any shift in the frequency of the PA line (within the experimental resolution) with increasing intensity.

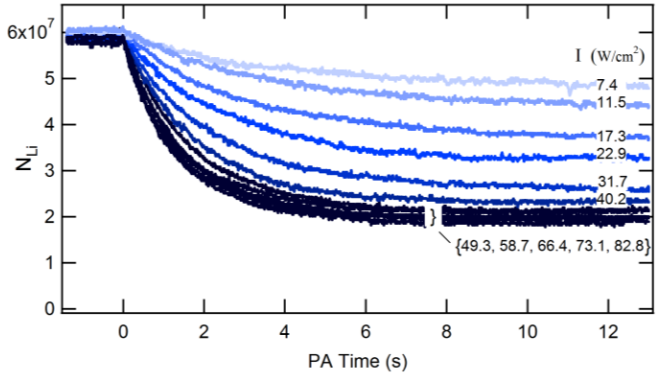

S4. Scheme for efficient production of LiRb molecules in the ground $X\,^1\Sigma^+\,(v''=0)$ state

The efficient production of ground $X\,^1\Sigma^+\,(v''=0)$ state molecules requires the following properties of the photo-associated state: (*i*) the outer turning point should be at a large inter-nuclear separation and (*ii*) the inner turning point should be at the same inter-nuclear distance as the equilibrium inter-nuclear distance $R_{eq}$ (= 6.55 $a_0$) of the $X\,^1\Sigma^+\,(v''=0)$ state. Unfortunately, in LiRb these two requirements are not simultaneously fulfilled by any of the observed PA levels. For example, the observed $v = 3$ level of the 4(1) state has outer turning point $R_{out}$ at ~ 34.1 $a_0$ (estimated based on the measured $C_6$ coefficient) and the inner turning point $R_{in}$ is ~ 5.8 $a_0$ (estimated from theoretical potential energy curves of Ref. 33). A vertically downward transition at $R_{in}$ would create $X\,^1\Sigma^+\,(v'' \sim 4)$ molecules which also have $R_{in} \sim 5.8\,a_0$ (primes and double primes indicate $v$ measured from the bottom of the respective potentials). The actual branching ratio can be measured using REMPI and, in fact, from our previous spectroscopy work [38], we find that the $B\,^1\Pi$ ($v' = 13$) state is a promising intermediate state for the detection of LiRb molecules produced in the $X\,^1\Sigma^+\,(v'' \sim 4)$ state.

In order to find the most efficient pathway for the creation of ground $X\,^1\Sigma^+\,(v''=0)$ state molecules it is instructive to compare the product ($\eta$) of the following two quantities: (*i*) the strength of the PA transition, which is determined by the free-bound Franck-Condon (FC) overlap between the wave function of the scattering state and the wave function of the PA level, and (*ii*) the strength of spontaneous emission from the PA level to different vibrational levels ($v''$) of the ground state, which is determined by the bound-bound FC overlap. Theoretical calculations of FC overlap using accurate potential energy curves indicate that the value of $\eta$ is quite high for loosely bound PA levels [40]. This indicates that photoassociating to a loosely bound level followed by spontaneous emission leads to high rate of formation of ground $X\,^1\Sigma^+$ state molecules. However, very few molecules are produced in the $v'' = 0$ level of the $X\,^1\Sigma^+$ state.

In order to roughly estimate the ground state molecule formation rate let's choose PA to $v = 3$ level of the 4(1) state which has a binding energy of around -42 GHz. For this state, the LiRb* formation rate is experimentally determined to be ~2×10$^7$ s$^{-1}$. Some of these LiRb* molecules would dissociate to form free Li and Rb atoms but some would spontaneously emit a photon to form LiRb molecules in the electronic ground $X\,^1\Sigma^+$ state (indeed, we have observed these electronic ground state molecules using REMPI). Assuming that ~20% of the LiRb* molecules spontaneously emit to form molecules in different $v''$ levels of the ground $X\,^1\Sigma^+$ state, we estimate that ~ 4×10$^6$ s$^{-1}$ ground $X\,^1\Sigma^+$ state (this assumption is the largest uncertainty in the estimate) molecules are formed. Among these, around 45% would populate the $v'' = 52$ level and around 0.35% would populate the $v'' = 4$ level (calculated from FC overlaps), leading to formation of ~ 1.8×10$^6$ and ~ 1.4×10$^4$ molecules per second in the $v'' = 52$ and $v'' = 4$ levels respectively.

The molecules in $v'' = 4$ level can be transferred to $v'' = 0$ level by exciting the molecules in the $v'' = 4$ level to the $v' = 2$ of the $B\,^1\Pi$ state. Around 25% of the molecules in the $v' = 2$ level of the $B\,^1\Pi$ state would decay to $v'' = 0$ level of the $X\,^1\Sigma^+$ state [38], producing ~ 3.5×10$^3$ molecules per second in the $v'' = 0$ level of the $X\,^1\Sigma^+$ state. The estimate is made assuming that only ~5×10$^7$ Li/Rb atoms were trapped in the MOT. It is relatively easy to increase the number of trapped atoms in the MOT by a factor of ten (we have done this in our apparatus too). This would increase the ground $X\,^1\Sigma^+$ ($v'' = 0$) state molecule formation rate to 3.5×10$^4$ s$^{-1}$, which is extremely promising for future experiments.

The overall efficiency is much higher if the LiRb molecules are transferred to the $X\,^1\Sigma^+\,(v'' = 0)$ state by more sophisticated techniques like STIRAP [12]. It is especially impressive if the 1.8×10$^6$ molecules produced per second in the $v'' = 52$ level are transferred to the $v'' = 0$ level of the $X\,^1\Sigma^+$ state. Preliminary theoretical estimates indicate that this is certainly feasible (with the $v' = 20$ level of the $B\,^1\Pi$ state being used as the intermediate state for the state transfer). Even if the STIRAP efficiency is relatively low (~50%), we expect that ~10$^6$ molecules produced per second in the $v'' = 0$ level of the $X\,^1\Sigma^+$ state which is unprecedented among any bi-alkali molecules.

* sourav.dutta.mr@gmail.com
† yongchen@purdue.edu

[1] K.-K. Ni, S. Ospelkaus, M. H. G. de Miranda, A. Pe'er, B.Neyenhuis, J. J. Zirbel, S. Kotochigova, P. S. Julienne, D. S. Jin, and J. Ye, Science **322**, 231 (2008).
[2] S. Ospelkaus, K.-K. Ni, D. Wang, M. H. G. de Miranda, B. Neyenhuis, G. Quéméner, P. S. Julienne, J. L. Bohn, D. S. Jin, and J. Ye, Science **327**, 853 (2010).
[3] C.-H. Wu, J. W. Park, P. Ahmadi, S. Will, and M. W. Zwierlein, Phys. Rev. Lett. **109**, 085301 (2012).
[4] M.-S. Heo, T. T. Wang, C. A. Christensen, T. M. Rvachov, D. A. Cotta, J.-H. Choi, Y.-R. Lee, and W. Ketterle, Phys. Rev. A **86**, 021602(R) (2012).
[5] J. Deiglmayr, A. Grochola, M. Repp, K. Mörtlbauer, C. Glück, J. Lange, O. Dulieu, R. Wester, and M. Weidemüller, Phys. Rev. Lett. **101**, 133004 (2008).
[6] J. M. Sage, S. Sainis, T. Bergeman, and D. DeMille, Phys. Rev. Lett. **94**, 203001 (2005).
[7] P. Zabawa, A. Wakim, M. Haruza, and N. P. Bigelow, Phys. Rev. A **84**, 061401(R) (2011).
[8] D. Wang, J. Qi, M. F. Stone, O. Nikolayeva, H. Wang, B. Hattaway, S. D. Gensemer, P. L. Gould, E. E. Eyler, and W. C. Stwalley, Phys. Rev. Lett. **93**, 243005 (2004).
[9] A. J. Kerman, J. M. Sage, S. Sainis, T. Bergeman, and D. DeMille, Phys. Rev. Lett. **92**, 033004 (2004).

[10] A. Ridinger, S. Chaudhuri, T. Salez, D. R. Fernandes, N. Bouloufa, O. Dulieu, C. Salomon, and F. Chevy, EPL (Europhys. Lett.) **96**, 33001 (2011).
[11] J. Ulmanis, J. Deiglmayr, M. Repp, R. Wester, and M. Weidemüller, Chem. Rev. **112**, 4890 (2012).
[12] K. Aikawa, D. Akamatsu, M. Hayashi, K. Oasa, J. Kobayashi, P. Naidon, T. Kishimoto, M. Ueda, and S. Inouye, Phys. Rev. Lett. **105**, 203001 (2010).
[13] D. DeMille, Phys. Rev. Lett. **88**, 067901 (2002).
[14] E. R. Hudson, H. J. Lewandowski, B. C. Sawyer, and J. Ye, Phys. Rev. Lett. **96**, 143004 (2006).
[15] M. A. Baranov, M. Dalmonte, G. Pupillo, and P. Zoller, Chem. Rev., **112**, 5012 (2012).
[16] M. Aymar and O. Dulieu, J. Chem. Phys. **122**, 204302 (2005).
[17] R. V. Krems, W. C. Stwalley, and B. Friedrich, COLD MOLECULES: Theory, Experiment, Applications. CRC Press, Taylor & Francis Group, 2009.
[18] S. Dutta, D. S. Elliott, and Y. P. Chen, EPL (Europhys. Lett.) **104**, 63001 (2013).
[19] H. Wang and W. C. Stwalley, J. Chem. Phys. **108**, 5767 (1998).
[20] J. L. Bohn and P. S. Julienne, Phys. Rev. A **60**, 414 (1999).
[21] I. D. Prodan, M. Pichler, M. Junker, R. G. Hulet, and J. L. Bohn, Phys. Rev. Lett. **91**, 080402 (2003).
[22] C. McKenzie, J. Hecker Denschlag, H. Häffner, A. Browaeys, L. E. E. de Araujo, F. K. Fatemi, K. M. Jones, J. E. Simsarian, D. Cho, A. Simonia, E. Tiesinga, P. S. Julienne, K. Helmerson, P. D. Lett, S. L. Rolston, and W. D. Phillips, Phys. Rev. Lett. **88**, 120403 (2002).
[23] U. Schlöder, C. Silber, T. Deuschle, and C. Zimmermann, Phys. Rev. A **66**, 061403(R) (2002).
[24] P. Pellegrini and R. Côté, New J. Phys. **11**, 055047 (2009).
[25] M. Yan, B. J. DeSalvo, Y. Huang, P. Naidon, and T. C. Killian, Phys. Rev. Lett. **111**, 150402 (2013).
[26] S. Dutta, A. Altaf, J. Lorenz, D. S. Elliott, and Y. P. Chen, J. Phys. B: At. Mol. Opt. Phys. **47**, 105301 (2014)
[27] M. Movre and R. Beuc, Phys. Rev. A **31**, 2957 (1985).
[28] B. Bussery, Y. Achkar, and M. Aubert-Frécon, Chem. Phys. **116**, 319 (1987).
[29] M. Marinescu and H. R. Sadeghpour, Phys. Rev. A **59**, 390 (1999).
[30] G. Herzberg, Molecular Spectra and Molecular Structure I: Spectra of Diatomic Molecules, D. Van Nostrand Company, Inc., Princeton, New Jersey (1961).
[31] K. M. Jones, E. Tiesinga, P. D. Lett, and P. S. Julienne, Rev. Mod. Phys. **78**, 483 (2006).
[32] W. C. Stwalley, M. Bellos, R. Carollo, J. Banerjee, and M. Bermudez, Mol. Phys. **110**, 1739 (2012).
[33] M. Korek, G. Younes, and S. AL-Shawa, J. Mol. Struct.: THEOCHEM **899**, 25 (2009).
[34] A. Grochola, A. Pashov, J. Deiglmayr, M. Repp, E. Tiemann, R. Wester, and M. Weidemüller, J. Chem. Phys. **131**, 054304 (2009).
[35] R. J. LeRoy and R. B. Bernstein, J. Chem. Phys. **52**, 3869 (1970).
[36] J. Deiglmayr, P. Pellegrini, A. Grochola, M. Repp, R. Côté, O. Dulieu, R. Wester, and M. Weidemüller, New J. Phys. **11**, 055034 (2009).
[37] C. Haimberger, J. Kleinert, O. Dulieu, and N. P. Bigelow, J. Phys. B: At. Mol. Opt. Phys. **39**, S957 (2006).
[38] S. Dutta, A. Altaf, D. S. Elliott, and Y. P. Chen, Chem. Phys. Lett. **511**, 7 (2011).
[39] I. Manai, R. Horchani, H. Lignier, P. Pillet, D. Comparat, A. Fioretti, and M. Allegrini, Phys. Rev. Lett. **109**, 183001 (2012).
[40] Jesús Pérez-Ríos, private communication.